\documentstyle[prl,twocolumn,aps,epsfig,amssymb]{revtex}
\def\beq{\begin{equation}}
\def\eeq{\end{equation}}
\def\bea{\begin{eqnarray}}
\def\eea{\end{eqnarray}}
\def\nnb{\nonumber}

\newcommand{\gsim}{\lower.7ex\hbox{$\;\stackrel{\textstyle>}{\sim}\;$}}
\newcommand{\lsim}{\lower.7ex\hbox{$\;\stackrel{\textstyle<}{\sim}\;$}}

\begin{document}
\twocolumn[\hsize\textwidth\columnwidth\hsize\csname@twocolumnfalse\endcsname

\title{ \vspace{-3ex}{ \small IC/2002/8; UPR-980-T
                      \hfill hep-ph/0202090 }\\[2mm]
Weak Mixing Angle and the $SU(3)_C\times SU(3)$ Model on 
$M^4\times S^1/(Z_2\times Z_2')$}

\vspace{0.2cm}
\author{	
Tianjun Li$^a$ and Wei Liao$^b$\\ }
\vspace{0.2cm}
\address{
        $^a$ Department of Physics and Astronomy, 
University of Pennsylvania, Philadelphia, PA 19104-6396, USA \\
E-mail: tli@bokchoy.hep.upenn.edu,
phone: (215) 573-5820, fax: (215) 898-2010\\
        $^b$ ICTP, Strada Costiera 11, 34014 Trieste, Italy
}
\maketitle
\begin{abstract}
We show that the desirable weak mixing angle $\sin^2\theta_W=0.2312$
at $m_Z$ scale can be generated naturally in the $SU(3)_C\times SU(3)$ 
model on $M^4\times S^1/(Z_2\times Z_2')$ where the gauge symmetry
$SU(3)$ is broken down to $SU(2)_L\times U(1)_Y$ by orbifold projection. 
For a supersymmetric model with a TeV scale extra dimension,
the $SU(3)$ unification scale is about hundreds of TeVs
at which the gauge couplings for $SU(3)_C$ and $SU(3)$
can also be equal in the mean time. For
the non-supersymmetric
model, $SU(2)_L \times U(1)_Y$ are unified at order of 10 TeV.
These models may serve as 
good candidates for physics beyond the SM or MSSM.
\vskip 0.2cm
\end{abstract}
]

Grand Unification Theory (GUT) has long been an intriguing proposal
for physics beyond the Standard Model (SM) or 
Minimal Supersymmetric Standard Model (MSSM)\cite{ggqw}.
The impressive successes of supersymmetric (SUSY) GUTs include the
successful prediction of the weak mixing angle at the electroweak (EW)
scale, the unification of three running gauge
couplings of the $G_{SM}=SU(3)_C\times SU(2)_L \times U(1)_Y$ group at GUT
scale $M_{GUT} \approx 2\times 10^{16}$ GeV. Besides the proton decay 
problem, an unsatisfactory feature of this idea
is that there exists a SUSY desert for new physics
from TeV scale to the GUT scale.
Several years ago, gauge coupling unification has been reconsidered in the
large extra dimension scenarios. The point of this proposal is that
beyond the compactification scale, the KK modes arising from the bulk fields
would give radiative corrections to
the gauge couplings which would behave 
as polynomial functions of the energy scale~\cite{ddg}. Gauge coupling
unification can be qualitatively
achieved around $10^6$ GeV if the compactification scale is
$10^5$ GeV, and the power law unification can be correct at quantitative level
if the high energy fundamental theory is known.

On the other hand, 
the discrete symmetry on the extra space manifold gives 
another interesting proposal for GUT breaking or gauge
symmetry breaking~\cite{alot,AHJM,SSSB,tj}.
The doublet-triplet splitting problem and proton decay problem
can be solved neatly in this kind of scenario. 
In addition, the discrete symmetry may not act freely on
the extra space manifold. When the discrete symmetry 
does not act freely on the extra space manifold, there exists a brane at 
each fixed point, line or hypersurface, where only part of 
the gauge symmetry and SUSY might be preserved and
the SM fermions can be located at~\cite{AHJM,tj}. However,
the SUSY desert still remains since the compactification
scale is close to the 4-dimensional (4D) GUT scale.

About thirty years ago, Weinberg tried to unify the $SU(2)_L$ weak and
$U(1)_Y$ hypercharge interactions into the $SU(3)$ gauge interaction
where the tree-level weak mixing angle ($\sin^2\theta_W$) arising from the breaking
 of a $SU(3)$ group is $0.25$ which is within $10\%$ of the
present experimental value of $0.2312$~\cite{SW}. However, the quark doublets
 can not be accommodated in the theory due to their small hypercharge
quantum numbers.
In a recent paper~\cite{dk}, Dimopoulos and Kaplan pointed out that
the above tree-level prediction of weak mixing angle ($\sin^2\theta_W=0.25$)
 can be obtained
even if the weak and hyperecharge interactions were not unified. 
And the new physics scale is expected to be around $3.75$ TeV using the SM gauge
couplings and beta functions. The idea is quite interesting but
the model they proposed is too complicated. Their model involved
the $SU(3)\times SU(2)\times U(1)$ gauge group altogether to get the SM
$SU(2)_L\times U(1)_Y$ gauge group by Higgs mechanism. 
One problem is that, as discussed in their paper, to achieve a good
value of $\sin^2\theta_W$ around $0.25$, the gauge couplings of the extra 
$SU(2)\times U(1)$ have to be much larger than that of $SU(3)$ group, which
determines the gauge couplings of the $SU(2)_L\times U(1)_Y$ group.
Consequently the $U(1)$ gauge group may have Landau pole problem
not far beyond the $3.75$ TeV. 

How to combine above ideas to construct a minimal simple model is 
an interesting
question. As we know, the gauge symmetry breaking by the discrete
symmetry on extra space manifold will not reduce the rank of
the bulk gauge group if the discrete symmetry is abelian. So,
only the $SU(5)$, $SU(4)_C\times SU(2)_L$ and $SU(3)_C\times SU(3)$ 
gauge symmetries
can be broken directly down to $G_{SM}$ without
extra U(1). The $SU(5)$ model
has been discussed extensively and the $SU(4)_C\times SU(2)_L$ have
relatively higher unification scale comparing to TeV.
In this letter, we would like to discuss the $SU(3)_C\times SU(3)$ model
on the space-time $M^4\times S^1/(Z_2\times Z_2')$ in which
the correct weak mixing angle $\sin^2\theta_W=
0.2312$ at the $m_Z$ scale can be obtained naturally with
a TeV scale extra dimension. By the way, the idea that part 
of the Standard Model particles live in large (TeV) extra dimensions
and the orbifold models based on $SU(3)\times SU(3)$ were also
discussed in Ref.~\cite{IA}.

We will concentrate on the SUSY
model because the discussion of non-SUSY model
is similar. Here is our convention. We consider
the 5-dimensional (5D) space-time as the product of the 
4D Minkowski space-time $M^4$, and the circle
$S^1$ with radius $R$. The corresponding
coordinates for the space-time are $x^{\mu}$, ($\mu = 0, 1, 2, 3$),
$y\equiv x^5$. In addition,
the orbifold $S^1/(Z_2\times Z_2^\prime)$ is obtained from
$S^1$ moduled by two equivalent classes: $y \sim -y$ and 
$y^\prime \sim - y^\prime$ where $y^\prime = y+ \pi R/2$.
$y=0, \pi R/2$ are two fixed points.
We assume that the bulk gauge symmetry is $G=SU(3)_C\times SU(3)$, and 
two Higgs hypermultiplets $\Psi_u$ and
$\Psi_d$ transforming as $(1,3)$ and $(1,\bar{3})$
are in the bulk. 

The $N=1$ SUSY theory with gauge group $G$
 in 5-dimension have 8 real supercharges,
corresponding to $N=2$ SUSY in 4-dimension. The vector multiplet
physically contains a vector boson $A_M$ where $M=0, 1, 2, 3, 5$, 
two Weyl gauginos $\lambda_{1,2}$, and a real scalar $\sigma$. 
In the language of 4D $N=1$ SUSY
, it contains a vector multiplet $V(A_{\mu}, \lambda_1)$ and
a chiral multiplet $\Sigma((\sigma+iA_5)/\sqrt 2, \lambda_2)$ in the
adjoint representation.
And the 5-dimensional hypermultiplet physically has two complex scalars
$\phi$ and $\phi^c$, two Weyl fermions $\psi$ and $\psi^c$, and can be 
decomposed into two chiral mupltiplets $\Phi(\phi, \psi)$
and $\Phi^c(\phi^c, \psi^c)$, which transform as conjugate 
representations of each other under $G$.

We define the $P$ and $P'$ as corresponding operators for 
the $Z_2$ symmetry 
transformations $y \to -y$ and $y^\prime \to - y^\prime$, respectively.
With general action for the 5-dimensional $N=1$ SUSY
gauge theory and its couplings to the bulk hypermultiplets in \cite{agw},
we obtain the symmetry transformations for the gauge fields,
and $\Phi_u$, $\Phi_u^c$ under $P$ 
\bea
&& V(x^\mu,y) \to V(x^\mu,-y) = P V(x^\mu,y) P^{-1}, \\
&& \Sigma(x^\mu,y) \to \Sigma(x^\mu,-y) = -P \Sigma(x^\mu,y) P^{-1}, \\
&& \Phi_u(x^\mu,y) \to \Phi_u(x^\mu,-y) = P \Phi_u(x^\mu,y), \\
&& \Phi^c_u(x^\mu,y) \to \Phi^c_u(x^\mu,-y) = -P^* \Phi_u^c(x^\mu,y), 
\eea
where $P^*$ is the complex conjugate of $P$.
And under $P$,
the transformation properties of $\Phi_d$ and $\Phi_d^c$ are
similar to those of $\Phi_u$ and $\Phi_u^c$.
Similar results hold for $P'$.
Because of the $Z_2$ symmetry,
the eigenvalues of $P$ and $P^\prime$ must be
 $\pm 1$. Denoting the general bulk field
$\phi$ with parities ($\pm, \pm$) under ($P$, $P'$) as
$\phi_{\pm \pm}$, the 
solutions under Fourier-expansion can be found in~\cite{bhn}.
Reducing to 4D fields, $\phi_{++}$ and $\phi_{--}$ 
correspond respectively to the KK modes with masses $2n/R$
and $(2n+2)/R$,  $\phi_{+-}$ and $\phi_{-+}$ are the KK modes
with masses $(2n+1)/R$ (n is non-negative integer).
We emphasize that the zero modes are contained only in
$\phi_{++}$ fields. $\phi_{--}$ and $\phi_{-+}$ would vanish
at $y=0$ and $\phi_{--}$ and $\phi_{+-}$ would vanish at $y=\pi R/2$.

We choose the matrix representaions for $P$ and $P'$ as
$P={\rm diag}(1,1,1)\otimes {\rm diag}(1,1,-1)$ and
$P^\prime={\rm diag}(1,1,1)\otimes {\rm diag}(1,1,1)$,
which are the group elements of $SU(3)_C\times SU(3)$ 
up to a $Z_2$ phase.
With general discussions and results for the gauge symmetry
breaking by the discrete symmetry on extra space manifold
 in~\cite{tj}, we obtain that
under $P$, the $SU(3)_C\times SU(3)$ 
gauge symmetry is broken down to
 $G_{SM}$
 on the 3-brane at $y=0$ for all the KK modes and in the bulk
for zero modes. And under $P'$, the 4D
$N=2$ SUSY is broken down to $N=1$ for zero modes.
The 3-branes at $y=0$ and $y=\pi R/2$ preserve only 4D
$N=1$ SUSY due to the $Z_2$ projections. 
These results can also be obtained by noticing
the parities of $V$ and
$\Sigma$ for the $SU(3)$ group, and of $\Phi_u$ and $\Phi_u^c$
(similar for $\Phi_d$ and $\Phi_d^c$) under $(P, P^\prime)$:
\bea
& V : \pmatrix{(+,+) & (+,+) & (-,+) \cr (+,+) & (+,+) & (-,+) \cr
(-,+) & (-,+) & (+,+) }, \\
& \Sigma : \pmatrix{(-,-) & (-,-) & (+,-) \cr (-,-) & (-,-) & (+,-) \cr
(+,-) & (+,-) & (-,-) }, \\
& \Phi_u : \pmatrix{(+,+) \cr (+,+) \cr (-,+)}, ~~
\Phi_u^c : \pmatrix{(-,-) \cr (-,-) \cr (+,-)}.
\eea
For $SU(3)_C$ group, $V$ and $\Sigma$
have parities $(+,+)$ and $(-,-)$, respectively.
Then, the MSSM quarks and leptons can be put on the observable 
3-brane at $y=0$, and the Higgs doublets $H_u$ and $H_d$
are obtained from the zero modes
of $\Phi_u$ and $\Phi_d$.
The Yukawa terms
are similar to those in the MSSM for on the observable 3-brane,
we only have Higgs doublets $H_u$ and $H_d$
for all the KK modes.

$SU(2)_L$ group is generated by $T^a \approx \sigma^a/2$ ($a=1,2,3$)
of the corresponding $SU(3)$ group,
where $\sigma^a$ is the Pauli matrix. $U(1)_Y$ is generated by 
$\sqrt{3} T^8$, so, the hypercharges of the remaining two Higgs
doublets $H_u$ and $H_d$ are $1/2$ and $-1/2$, respectively.
From the $SU(3)$ relation we would get $g_1=g_2/\sqrt{3}$.
On the observable 3-brane at $y=0$, we can then put the left handed quark 
and lepton fields in the fundamental representation of the $SU(2)_L$
group while the right handed quarks and leptons are singlets.
Knowing that the total $U(1)_Y$ charges for each Yukawa term must be zero,
together with four anomaly-free conditions: $(3_C)^2 Y$, $(2_L)^2 Y$,
$Y^3$ and $(grav)^2 Y$, we can determine the correct hypercharges
for $Q$, $U$, $D$, $L$ and $E$ fields. Since the hypercharges of
two Higgs doublets are quantized as they are from triplets of $SU(3)$ 
group, an understanding of the charge quantization is achieved as a 
result of the consistency of our setup.

There exist two energy scales in our models: the 
compactification scale $M_C=1/R$, and the $SU(3)$ unification
scale which is considered as
the cutoff scale $\Lambda$ in the theory. 
Because there might exist additional
 $SU(2)_L$ and $U(1)_Y$ kinetic terms localized on the observable
3-brane which may violate the $SU(3)$ gauge coupling relation,
 the cutoff scale might be much larger than the compactification scale
if those localized kinetic terms were not small. 
We will neglect those extra kinetic terms for two reasons:
(1) It is natural to set those localized kinetic terms to zero at tree
level in the fundamental theory, so, they can only be generated at
loop level as counter terms and are very small if the theory
is weakly coupled at cutoff scale; (2) If the gauge interaction
is strong coupled at $\Lambda$ and $M_C\sim 0.01 \Lambda$, the
effects of those kinetic terms are very small~\cite{hall}.

First, let us study the scenario with $M_C\sim \Lambda$.
Assuming the sparticle mass
scale, $M_S$, in the range $200-1000$ GeV, using the MSSM beta function 
$(-11,-1, 3)$ above this scale and the SM beta function $(-41/6,19/6,7)$ 
down to the $m_Z$ scale, in the limit that $\sin^2\theta_W=0.25$ is realised
at $M_C$ or $\Lambda$, we find out that $M_C \sim 77-14$ TeV.

Second, we discuss the scenario with $M_C \sim 0.01 \Lambda$. 
For $M_C < \mu < \Lambda$, we should include the massive KK modes in
counting the radiative corrections to gauge couplings.
There are two kinds of beta functions, $b_o$ and $b_e$, from the
KK modes of bulk fields with masses $2n/R$ and $(2n+1)/R$, respectively.
The radiative corrected gauge couplings above the $M_C$ scale is
\bea
     \alpha_i^{-1}(\mu) &=&
      \alpha_i^{-1}(M_C)
  + {b_i \over 2\pi} \,\ln{\mu \over M_C} \nnb \\
 && + \sum_{n} \frac{1}{2 \pi} ( b_{io} \ln{\mu \over M_{2n-1}}
 +b_{ie} \ln{\mu \over M_{2n}}),
\label{run}
\eea
where $\alpha_i=g^2_i/4\pi$, $b_e= (-2,2,6)$ and $b_o=(14,2,0)$.
One obtains that the compactification scale is several TeVs in this
kind of scenario.

Third, because we include the 
$SU(3)_C$ gauge symmetry in the bulk, it is interesting to
study whether there exists the scenario in which
 the $SU(3)$ unification scale or $\Lambda$ is the
scale at which the gauge couplings for
$SU(3)_C$ and $SU(3)$ are also equal in the mean time.
And if the 5D $SU(3)$ theory is strong coupled at cutoff scale,
it is reasonable to assume that the gauge coupling for
$SU(3)$ is similar to that of $SU(3)_C$ at $\Lambda$.
We would like to emphasize that because we do not
present an explicit Grand Unified Theory which
 unifies the $SU(3)$ and $SU(3)_C$ gauge interactions in this letter,
we do not mean that there exists
 the gauge unification for $SU(3)$ and $SU(3)_C$ even if
 the gauge couplings for $SU(3)_C$ and $SU(3)$ are equal or similar
at the cut-off scale. 
However, the quantitative 
study can show whether a further complete gauge coupling
unification might be possible or not
from the point of view of gauge coupling runnings.

\begin{figure}[t]
\centerline{\psfig{figure=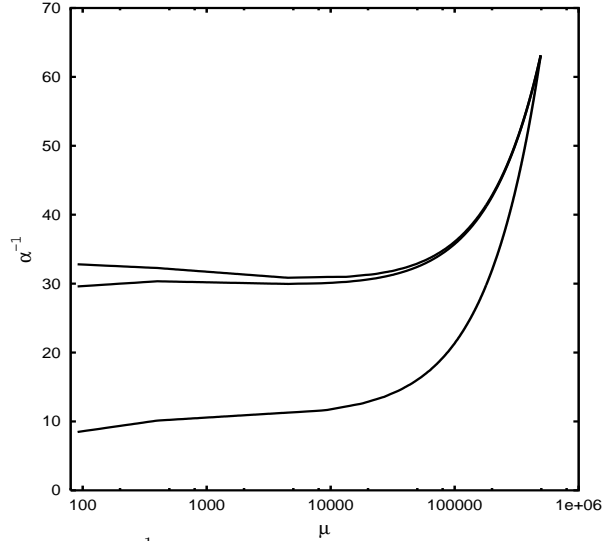,height=7cm,width=8cm}}
\caption{\small $\alpha^{-1}$ versus energy scale. From below three lines are
for $\alpha_3^{-1}$, $\alpha_2^{-1}$ and $\alpha_1^{-1}/3$.
 }
\label{fig}
\end{figure}

To have an estimate, we extract the leading $SU(3)$ symmetric 
contributions given by the KK modes,
that is $\sum_k \frac{b_{io}+b_{ie}}{4 \pi} \ln{\mu \over M_k}$.
Similar to the case in \cite{ddg}, for $\mu \gg M_C$, we may resum
this leading contribution to be a power law running, that is
$\frac{b_{io}+b_{ie}}{4 \pi}(\frac{\mu}{M_C}-1)$. Using the beta
function listed above, we get a rough estimate of the cutoff scale $\Lambda$:
\bea
{1 \over 2 \pi}({\Lambda \over M_C}-1) \approx \alpha^{-1}_2-\alpha^{-1}_3,
\eea
where $\alpha_{2,3}$ are the running gauge couplings for $SU(2)_L$ and
$SU(3)_C$ at scale $M_C$. The difference in the power
law running is given by the KK tower from the Higgs fields which
we choose living in the bulk. ${\Lambda \over M_C}$ is about $100-130$ for
$M_S \sim 200-1000$ GeV and $M_C < 10$ TeV. At first sight this
large gap between the cutoff scale and $M_C$ 
requires less than TeV $M_C$ to meet the purpose of unifying simultaneously
$SU(2)_L$ and $U(1)_Y$, 
because the $SU(3)$ asymmetric relative running
given by the zero mode is $\frac{b_1/3-b_2}{2 \pi} ln\frac{\Lambda}{M_C}$
which is about $-2.03$ for $\Lambda/M_C =120$ ($\alpha^{-1}_1/3
-\alpha^{-1}_2 \approx 3.21$ at $m_Z$ scale). However, there
are also $SU(3)$ asymmetric radiative corrections
given by the massive KK modes. It is $\sum_k \frac{b_{o}
-b_e}{4 \pi} ln\frac{M_{2k}}{M_{2k-1}} =
\frac{b_o -b_e}{4 \pi} ln\frac{2}{1}{4 \over 3}{6 \over5}\cdots$.
This relative running between $3\alpha_1$ and $\alpha_2$ is completely
given by the $U(1)_Y$ part and is about $1.11$ for $\Lambda/M_C =120$.
Detailed calculation using logarithmic running including step by step
massive KK states shows that in the whole range $M_S \sim 200-1000$ GeV,
$3 \alpha_1$, $\alpha_2$ and $\alpha_3$ can meet simultaneously 
for $M_C \sim 9.3-1.5$ TeV with $\alpha^{-1}\sim 60-66$ and
$\Lambda/M_C$ around 110. 
And the requirement that $3 \alpha_1$, $\alpha_2$ and $\alpha_3$
 meet simultaneouly further constrains the model, {\it i. e.},
$M_C$ and $\Lambda$ will be correlated if the SUSY threshold is known.
 In Fig. \ref{fig}, we give
a plot for $M_S=400$ GeV showing the $\alpha^{-1}$ versus the energy
scale. We may see clearly the running behaviours of the three gauge
couplings. $3 \alpha_1$, $\alpha_2$ and $\alpha_3$
meet with a value of about $1/63$ at $\Lambda \approx 494$ TeV. 
In this case $M_C \approx 4.5$ TeV and $\Lambda/M_C \approx 110$.
We may see that for $\mu \gg 10$ TeV, the difference between 
$3 \alpha_1$ and $\alpha_2$ is quite small (the discrepancy is less 
than $0.2\%$ for $\mu > 100$ TeV).

The SUSY breaking can be done by the Scherk-Schwarz mechanism
or by gauge mediated SUSY breaking. For the first approach,
one can solve the $\mu$ problem, and get the gaugino masses,
the $\mu B$ terms and Higgs masses~\cite{SSSB}. 
The radiative electroweak (EW) symmetry breaking may
also be realized. For the second approach,
the 3-brane at the fixed point $y=\pi R/2$ can be considered as
hidden sector for SUSY breaking. The SUSY breaking
effects may be communicated to the observable sector via gauge-mediated
mechanism. This may be realized if there is a SUSY breaking gauge 
singlet field $S$ living in the bulk. Interaction like $Tr(S W^\alpha
W_\alpha)$ gives gaugino masses via $F$-term. $\mu$ term, $\mu B$ and Higgs
mass terms can also be generated via the interaction of $\Phi$ and 
$S$\cite{susyb}. Squarks and sleptons can then obtain their masses 
through the radiative corrections.
EW symmetry breaking may be achieved in the usual way
via loop corrections at the EW scale because of the large
top quark Yukawa coupling\cite{wei}.

Similarly, we can construct a non-SUSY model
on $M^4\times S^1/(Z_2\times Z_2')$.
The triplet scalar Higgs field can live in the bulk, too. The SM fermions
are confined on the observable 3-brane. 
Because the relative running between 
$3\alpha_1$ and $\alpha_2$ are much faster than that in the SUSY case
for $\mu < M_C$, and the relative power law running between $3\alpha_1$
and $\alpha_3$ or between $\alpha_2$ and $\alpha_3$ is $1/12$ times
slower for $\mu > M_C$, 
it is impossible that the three gauge couplings meet simultaneously
at an energy scale if $M_C>1$ TeV involved.
So, it may be better to put $SU(3)_C$ on the boundary.
Since $\sin^2\theta_W=0.25$ 
is expected to be reached using the SM beta functions at about
$3.75$ TeV, we should have $M_C < 3.5$ TeV for this case. The cutoff scale,
which is understood as the energy scale where $3 \alpha_1$ and $\alpha_2$
unified, can be tens of TeV by setting $M_C > 1$ TeV.
The problem of this case is that
the masses of Higgs zero modes might be a problem.
Studying on EW symmetry breaking,
if it's accounted for solely by the radiative corrections (for which $1\%$
fine tuning is clearly needed), would give us the
strong constraints on this setup.

Furthermore, we can discuss the $SU(3)_C\times SU(3)$ model
on $M^4\times S^1/Z_2$ with gauge-Higgs unification. However,
the weak mixing angle $\sin^2\theta_W$ is expected about $0.75$ at the
cutoff scale.

The constraints on $M_C$ from the current high energy experiment
data are model dependent, and one need to study it in
detail. It is possible that the present knowledges might give
the strong constraints on the non-SUSY model for $M_C < $ 3.5 TeV.
Moreover, the extra $SU(2)_L$ doublets in the off-diagonal
$\Sigma$ ($A_5$ for non-SUSY case), and the third components
of the $\Phi_u^c$ and $\Phi_d^c$, may be produced
at LHC via s-channel $W^{\pm}$, $Z^0$ and $\gamma$ 
exchange for $M_C$ is several
TeVs, and the decay modes are model dependent. 
For the $SU(2)_L$ doublets in the off-diagonal 
$V$ ($A_{\mu}$ for non-SUSY case) and the
third components of the $\Phi_u$ and $\Phi_d$,
 there might exist the derivative interactions with the leptons,
so, they might also be produced at LHC and decay into leptons.
Of course, the decay channels of the KK modes of bulk fields are
dependent on the loop corrections to the masses of KK modes,
which should be considered in the collider test of the
models.

{\it Conclusion.} We have presented the $SU(3)_C\times SU(3)$ model
on the space-time $M^4\times S^1/(Z_2\times Z_2')$ with natural
explanation for the weak mixing angle which is a consequence of the
$SU(3)$ symmetry breaking in our setup. 
For the SUSY model with a TeV scale extra dimension,
the $SU(3)$ unification scale is about hundreds of TeVs
at which the gauge 
couplings for $SU(3)_C$ and $SU(3)$ can be equal in the mean time.
For the non-SUSY model, the
$SU(3)$ unification of $SU(2)_L \times U(1)_Y$ may be realized
at tens of TeVs at which the gauge couplings for
$SU(3)_C$ and $SU(3)$ can not be equal simultaneously.
Therefore, the supersymmetric desert does not appear in our models,
which is an unsatisfactory feature in the usual GUTs.
And our study shows that from the point of view of the gauge
coupling runnings, a complete gauge coupling
unification might occur for the SUSY case.
We stress that both models presented in this letter, supersymmetric or not, 
are of strong predictive power because of their clean setup.
Studying the non-supersymmetric model, one may
soon reveal whether it's realistic or not. 
For the supersymmetric model, we can break the SUSY by
Scherk-Schwarz mechanism or gauge mediated SUSY breaking,
and might test it at LHC.
We also comment on the charge quantization, 
radiative EW symmetry
breaking, and the models with gauge-Higgs unification. 
We conclude that this kind of interesting models may 
be good candidates for the minimal extensions of the SM or MSSM,
and their phenomenology deserves further study.

After sending our letter, we noticed the papers~\cite{hn,kdw} which discuss
similar issue.

{\it Acknowledgments.} We would like to thank 
W. Y. Keung for reading the manuscript and
useful discussions. This work was supported in part by
the U.S.~Department of Energy under Grant No.~DOE-EY-76-02-3071.


\begin{thebibliography}{99}
\bibitem{ggqw} H. Georgi and S. Glashow, Phys. Rev. Lett. {\bf 32} (1974) 438;
H. Georgi, H. Quinn and S. Weinberg, Phys. Rev. Lett. {\bf 33} (1974) 451;
 J. Pati and A. Salam, Phys. Rev. D{\bf 8} (1973) 1240,
Phys. Rev. D{\bf 10} (1974) 275;
S. Dimopoulos and H. Georgi, Nucl. Phys. B{\bf 193} (1981) 150.
\bibitem{ddg}K.R. Dienes, E. Dudas, and T. Gherghetta, Phys. Lett. {\bf
B436} (1998) 55, Nucl. Phys. {\bf B537} (1999) 47.
\bibitem{alot} Y. Kawamura, Prog. Theor. Phys. {\bf 103} (2000) 613,
Prog. Theor. Phys. {\bf 105} (2001) 999, Prog. Theor. Phys. {\bf 105} (2001) 691;
G. Altarelli and F. Feruglio, Phys. Lett. B{\bf 511} (2001) 257;
L. Hall and Y. Nomura, Phys. Rev. D{\bf 64} (2001) 055003.
\bibitem{AHJM} A. Hebecker and J. March-Russell, Nucl. Phys. B{\bf 613} (2001) 3.
\bibitem{SSSB} R. Barbieri, L. Hall and Y. Nomura, hep-ph/0106190.
\bibitem{tj} T. Li, hep-th/0112255.
\bibitem{SW} S. Weinberg, Phys. Rev. {\bf D5} (1972) 1962.
\bibitem{dk} S. Dimopoulos and D. E. Kaplan, hep-ph/0201148.
\bibitem{IA} I. Antoniadis, Phys. Lett. {\bf B246} (1990) 317;
I. Antoniadis and K. Benakli, Phys. Lett. {\bf B326} (1994) 69;
I. Antoniadis, K. Benakli and M. Quiros, hep-th/0108005. 
\bibitem{bhn} R. Barbieri, L. Hall and Y. Nomura, 
Phys. Rev. D{\bf 63} (2001) 105007.
\bibitem{agw} N. Arkani-Hamed, T. Gregoire and J. Wacker, hep-th/0101233.
\bibitem{hall} L. Hall and Y. Nomura, hep-ph/0111068.
\bibitem{susyb} D. E. Kaplan, et.al.,
Phys. Rev. D{\bf 62} (2000) 035010;
Z. Chacko et.al., JHEP{\bf 0001} (2000) 003.
\bibitem{wei} S. Weinberg, The Quantum Theory of Fields, Vol. III,
Cambridge University Press, 2000.
\bibitem{hn}  L. J. Hall and Y. Nomura, hep-ph/0202107.
\bibitem{kdw} S. Dimopoulos, D. E. Kaplan and N. Weiner, hep-ph/0202136.
\end{thebibliography}
\end{document}